\documentstyle[prl,aps,multicol,epsfig]{revtex}
\renewcommand{\narrowtext}{\begin{multicols}{2}
\global\columnwidth20.5pc} 
\renewcommand{\widetext}{\end{multicols}
\global\columnwidth42.5pc} \multicolsep = 8pt plus 4pt minus 3pt

\begin{document}
\draft
\title{A single-electron inverter}
\author{C. P. Heij, P. Hadley, and J. E. Mooij}
\address{Applied Physics and DIMES, Delft University of Technology\\
Lorentzweg 1, 2628 CJ Delft, The Netherlands}
\date{\today}
\maketitle

\begin{abstract}
A single-electron inverter was fabricated that switches from a high output 
to a low output when a fraction of an electron is added to the input. For the proper operation of the inverter, 
the two single-electron transistors that make up the inverter must exhibit voltage 
gain. Voltage gain was achieved by fabricating a combination of parallel-plate gate 
capacitors and small tunnel junctions in a two-layer circuit.
Voltage gain of 2.6 was attained at 25 mK and remained larger than one for
temperatures up to 140 mK. The temperature dependence of the gain agrees
with the orthodox theory of single-electron tunneling.
\end{abstract}


\narrowtext 
The use of single-electron tunneling devices for computation has been widely
discussed because these devices can be made very small and they
consume relatively little power. \cite{likharev99,pash00} A variety of
single-electron device logic schemes have been put forward but relatively
few of the proposed single-electron logic elements have been tested
experimentally. Here, measurements on a single-electron inverter are
presented. The inverter is a fundamental building block of single-electron
transistor logic, which bears considerable resemblance to standard
Complementary Metal-Oxide-Semiconductor (CMOS) logic. \cite{tucker92,korotkov95} The logic levels are
represented by voltages and a small number of electrons are transported when
the inverter switches from the high state to the low state. The logic gates
NAND and NOR can be realized by making slight variations on the inverter
circuit. With two inverters, a static Random Access Memory (RAM) memory cell can be constructed and
a ring oscillator can be made from three inverters. Voltage gain is essential 
for many of these circuit applications but thus far gain has been
achieved in relatively few devices. \cite{zimmer92,heij99} The inverter
discussed here has voltage gain.

A scanning electron microscope (SEM) photograph of the device and a
schematic diagram of the inverter circuit are shown in Fig. 1. The inverter
consists of two nominally identical single-electron transistors (SET's) in
series that share a common input gate. Each single-electron transistor used
in this inverter contains a small aluminum island with a total
capacitance $C_{\Sigma }=1.6$ fF. The SET's are outlined with dotted
lines in the schematic. The input of the inverter, $V_{i}$, extends under the two
islands. The input is electrically isolated from the two islands by an 8 nm
thick layer of aluminum oxide. This forms the two input capacitors $C_{i1}$ and 
$C_{i2}$. The output $V_{o}$ is connected to ground via an on-chip load capacitor 
$C_{L}$ of 130 pF to suppress charging effects at the output. The output is also
coupled to the two islands via small tunnel junctions. The power lead, $V_{b}$, 
and the grounded lead are similarly connected to the islands via small tunnel 
junctions. The two tuning gates, $V_{g1}$ and $V_{g2}$, are used

\begin{figure}[hbt]
\epsfig{figure=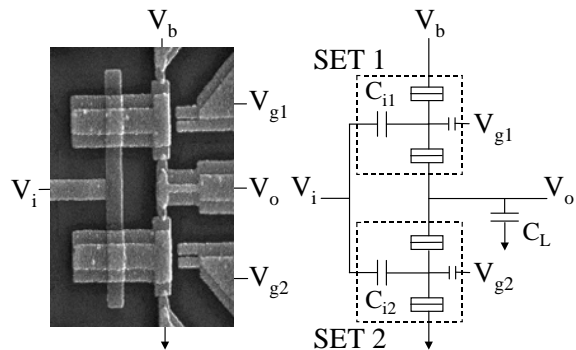, width=18pc, clip=true}
\caption{On the left is a scanning electron microscope image of the inverter
and on the right is the corresponding schematic. The width of the T-shaped
islands is 1 $\mu$m. The dotted lines outline the two SET's in the circuit. The load
capacitor $C_L$ is not shown in the SEM photo.}
\label{fig1}
\end{figure}

\noindent to tune the induced charges on the two islands.
For inverter operation, the output should be high when the input is low and 
the output should be low when the input is high. This is achieved by 
applying a bias voltage of $V_{b}=e/C_{\Sigma }=100\mu$V and adjusting the two 
tuning gates such that when 
the input voltage is low, the top transistor is conducting and the bottom 
transistor is in the Coulomb blockade. This effectively connects the output to the 
supply voltage and makes the output high. The Coulomb blockade prevents a steady 
current from flowing through the inverter limiting the dissipation in this device. 
When a high input voltage is applied, this shifts the induced charge on 
each of the SET's by a fraction of an electron and puts the top transistor in 
Coulomb blockade and makes the bottom transistor conducting so that the output 
is effectively connected to ground. Thus, when the input is high, the output
is low.

The device was fabricated on a thermally oxidized silicon substrate using a 
high-resolution electron beam pattern generator at 100 kV. Each layer of the 
circuit was defined using a double layer resist and was aligned to prefabricated 
Pt markers. The bottom layer of the circuit consisted of a 25 nm thick
aluminum film that was patterned to form the lower electrodes of the gate 
capacitors and the load capacitor. To form the dielectric for the capacitors, 
the sample was heated to 200$^{\circ }$C and the aluminum was oxidized in an 
O$_{2}$ plasma at 100 mTorr for 5 minutes. The resulting Al$_{x}$O$_{y}$ layer 
was 8 nm thick. A second aluminum film was then deposited in a pattern that 
defined the islands and the leads. The four tunnel junctions were defined in 
this layer by shadow evaporation.

The device was measured in a dilution refrigerator with a base temperature of 25 mK. 
All leads were equipped with $\pi $-filters at room temperature and copper-powder 
filters at base temperature. Superconductivity was suppressed using a 1 T magnetic field.
Measurements on the individual SET's revealed that the tunnel junctions were 
all identical with capacitances of $C_{j}=280$ aF and resistances of 1.1 
M$\Omega $. The input capacitances were $C_{i1}=800$ aF and $C_{i2}$=$810$ aF, 
the tuning gate capacitances were $45$ aF, and the stray capacitance of each island
was estimated to be $190$ aF. 

Figure 2a shows the dependence of the output voltage on
$V_i$ and $\Delta Q$, the difference between the induced charge on both islands.
The induced charge difference can be controlled by adjusting the two tuning gates $V_{g1}$
and $V_{g2}$. This figure illustrates that the
input-output characteristics of the inverter can vary greatly depending on
the voltages applied to the tuning gates. In the region around $\Delta Q=e$, 
the output is a weak function of the input. When the tuning gates are adjusted 
so that the induced charges are $e/2$ out of phase, there are large oscillations 
of the output voltage as the input voltage is varied.

\begin{figure}[hbt]
\epsfig{figure=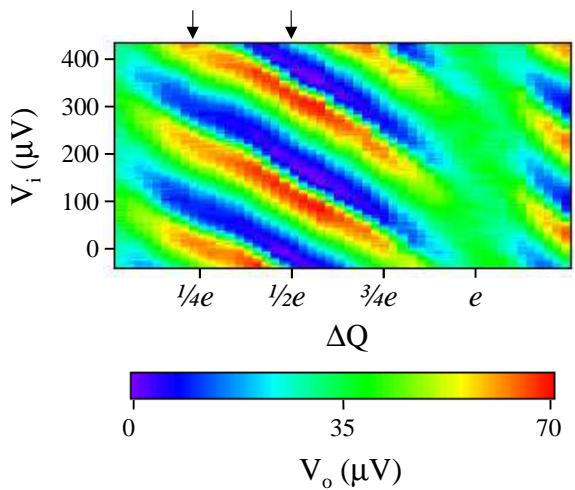, width=18pc, clip=true}
\caption{The output voltage of the inverter is plotted in color as a
function of the input voltage and the induced charge difference between the two islands.
The arrows indicate where the data for Fig. 3 was extracted from Fig. 2.}
\label{fig2}
\end{figure}

Increasing both of the tuning gates simultaneously leaves $\Delta Q$ constant but has the same effect as increasing $V_i$. 
Figure 3 shows input-output characteristics at 30 mK for two different
values of $\Delta Q$. The arrows in Fig. 2 indicate where the data for Fig. 3
was extracted. Note that when the input voltage is low, the output voltage
is high and when the input voltage is high, the output voltage is low. 
To achieve this input-output characteristic, the
gate voltages were adjusted manually. At this point there is no procedure
known for automatically tuning the gate voltages on-chip for the optimal
inverter performance. This is probably the largest problem
inhibiting the further development of this sort of logic.

The solid lines in Fig. 3 are simulations of the inverter characteristic
calculated at 30 mK using the orthodox theory for single-electron tunneling.
For the most part, the orthodox theory fits the measurements. The largest
deviations occur at $\Delta Q=e/2$ and either minimal or maximal output voltage. 
At $\Delta Q=e/2$, the maximal observed output voltage swing is 75\% of what is
expected from orthodox theory. This can be due to external noise coupling in
via the leads. The device is most sensitive to external noise at these bias conditions.

\begin{figure}[hbt]
\epsfig{figure=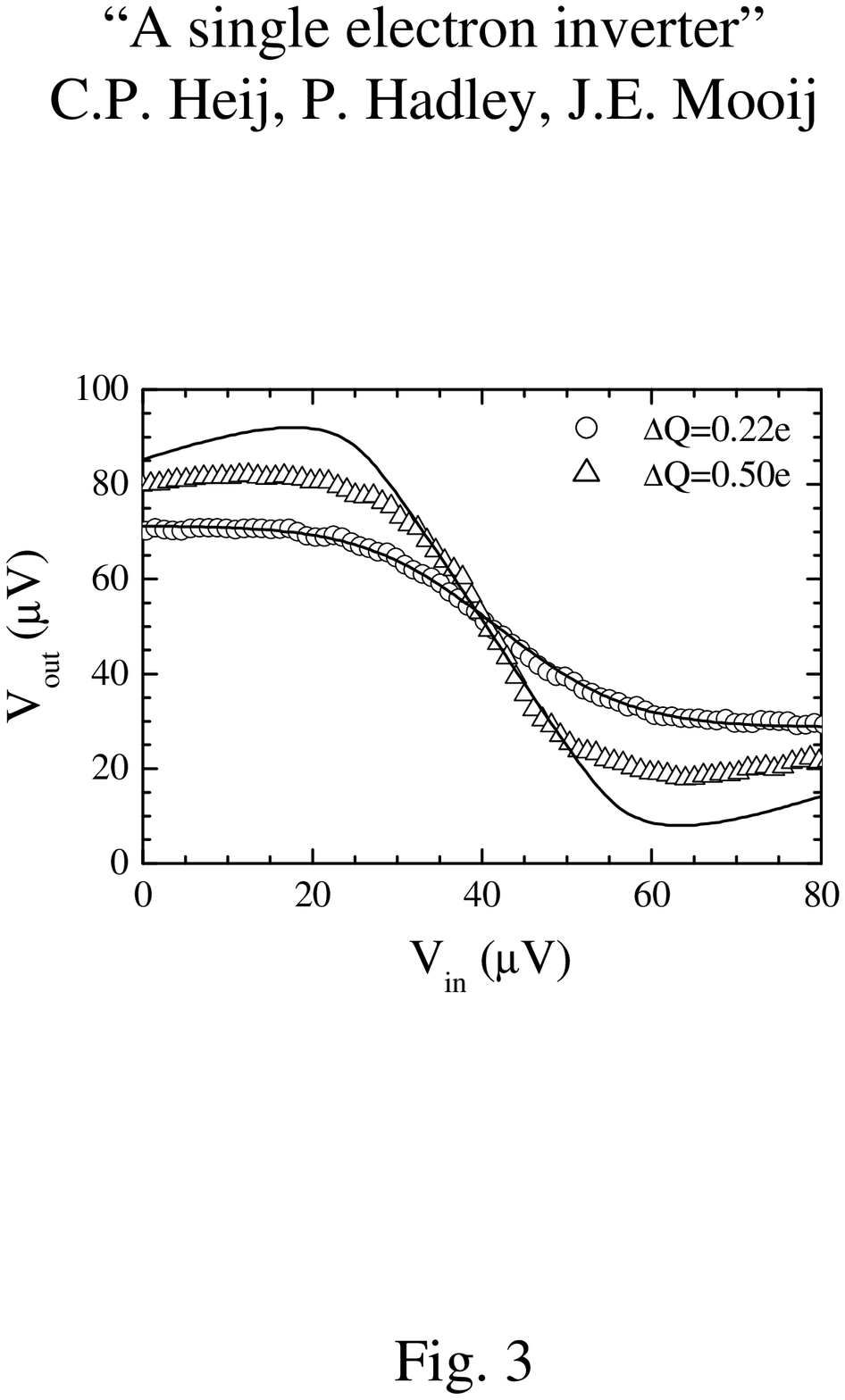, width=18pc, clip=true}
\caption{The input-output characteristics of the inverter are plotted for
two values of the induced charge difference. The solid lines are simulations using
the orthodox theory for single-electron tunneling.}
\label{fig3}
\end{figure}

An important consideration for the proper operation of an inverter is that
the maximum output voltage swing must be greater than the voltage swing
necessary at the input to switch the output from low to high. In other
words, the inverter must exhibit voltage gain. The maximum voltage gain that
can be achieved in a single-electron transistor is the ratio of the input-gate
capacitance to the junction capacitance $g_{v}=C_{i}/C_{j}$. In this work,
overlap capacitors were used to make relatively large input-gate capacitors.
The maximum voltage gain of the inverter, which can be determined from the
slope of the transitional region in Fig. 3, was $g_{v}=2.6$.

\begin{figure}[hbt]
\epsfig{figure=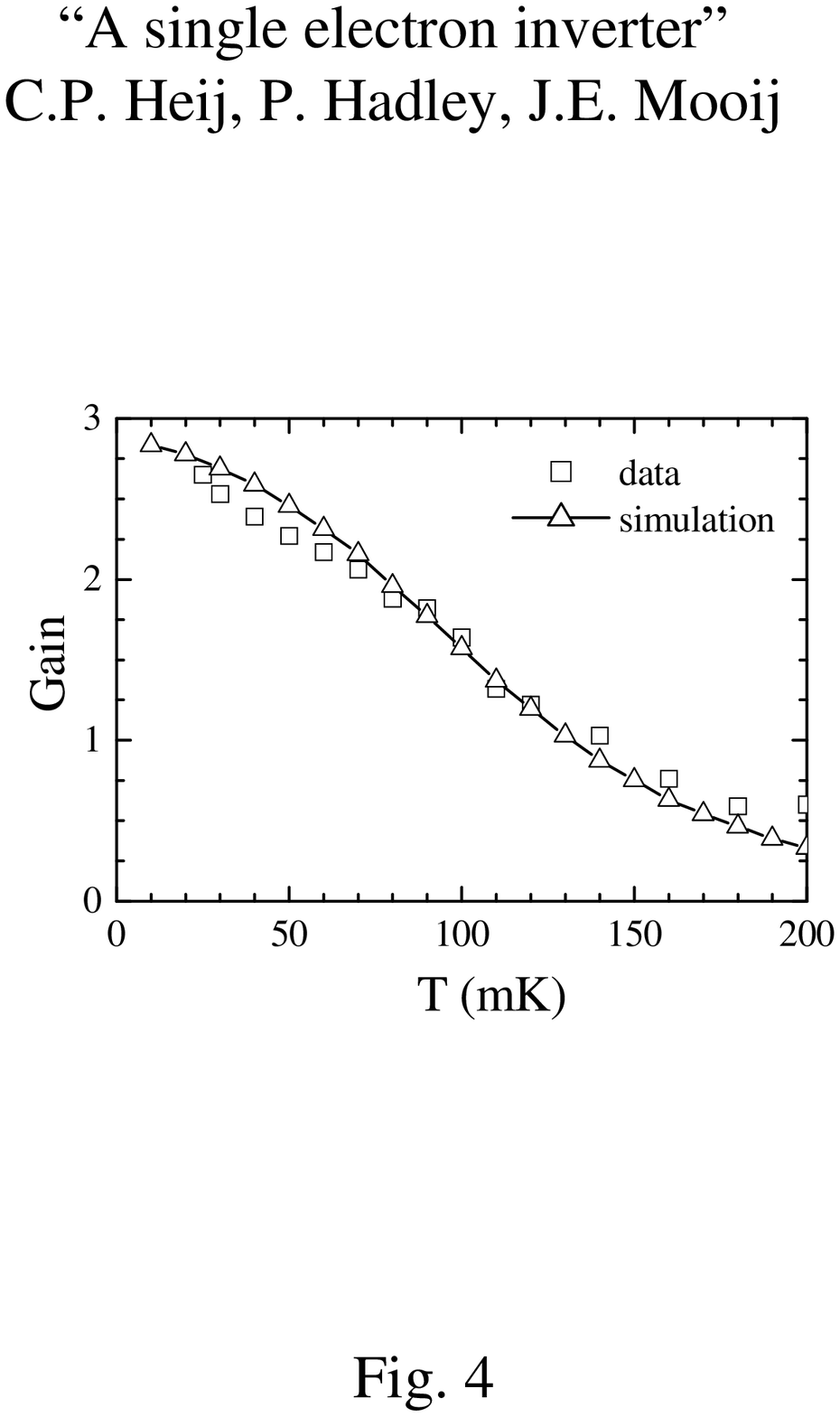, width=18pc, clip=true}
\caption{The gain of the inverter is plotted as a function of the
temperature. The gain decreases below one at 140 mK. The expected
temperature dependence of the gain was simulated using the orthodox theory
for single-electron tunneling.}
\label{fig4}
\end{figure}

Figure 4 shows the gain of the inverter as a function of temperature. Note
that a voltage gain greater than one is attained for temperatures below
about 140 mK. This is the highest temperature for which voltage gain in a
single-electron transistor has been achieved. While there have been many
reports of single-electron transistors operating at room temperature, those
transistors typically have a voltage gain much less than one. 
Gain is difficult to achieve at high temperatures because the gate capacitance 
must be made larger than the junction capacitance while making the total 
capacitance small. Fabricating a room-temperature single-electron transistor 
with voltage gain is challenging because it almost certainly requires control 
of the fabrication process on a nanometer scale in three dimensions.

The measurements on this single-electron inverter show that it can operate
as it was designed to and that the orthodox theory adequately describes the
behavior of the circuit. The voltage gain necessary for the operation of this
inverter was achieved by using overlap capacitors in a multilayer circuit, yielding
a maximum gain of 2.6 at 25 mK.

\widetext 

\end{document}